\documentclass[aps,prl,superscriptaddress,twocolumn]{revtex4-2}

\usepackage{amsfonts}
\usepackage{amsmath}
\usepackage{mathrsfs}
\usepackage{amssymb}
\usepackage{graphicx}
\usepackage{bm}
\usepackage{color}
\usepackage[colorlinks=true,linkcolor=blue,anchorcolor=blue,
citecolor=blue,urlcolor=blue]{hyperref}
\usepackage[normalem]{ulem}
\usepackage{float}

\newcommand{\R}{\mathrm{R}}

\begin{document}
\preprint{APS/123-QED}

\title{Lattice dynamics in an emergent Zeeman lattice}
\author{M. K. H. Ome}
\affiliation{Department of Physics and Astronomy, Washington State University, Pullman, Washington 99164-2814}
\author{Huaxin He}
\affiliation{Department of Physics, Shanghai University, Shanghai 200444, China}
\author{A. Mukhopadhyay}
\affiliation{Department of Physics and Astronomy, Washington State University, Pullman, Washington 99164-2814}
\author{E. Crowell}
\affiliation{Department of Physics and Astronomy, Washington State University, Pullman, Washington 99164-2814}
\author{S. Mossman}
\affiliation{Department of Physics and Astronomy, Washington State University, Pullman, Washington 99164-2814}
\author{T. Bersano}
\affiliation{Department of Physics and Astronomy, Washington State University, Pullman, Washington 99164-2814}
\author{Yongping  Zhang}
\email{yongping11@t.shu.edu.cn}
\affiliation{Department of Physics, Shanghai University, Shanghai 200444, China}
\author{P. Engels}
\email{engels@wsu.edu}
\affiliation{Department of Physics and Astronomy, Washington State University, Pullman, Washington 99164-2814}

\begin{abstract}
Periodic band structures are a hallmark phenomenon of condensed matter physics. While often imposed by external potentials, periodicity can also arise through the interplay of couplings that are not necessarily spatially periodic on their own. Here, we investigate dynamics in a lattice structure that emerges from the simultaneous application of Raman and radio frequency coupling to a dilute-gas Bose-Einstein condensate.  We demonstrate a variety of techniques including Kapitza-Dirac scattering, Bloch oscillations, and lattice shaking with spin and momentum resolved measurements.  
This combined coupling scheme allows for exceptional tunability and control, enabling future investigations into unconventional band structures such as quasi-flat ground bands and those with semimetal-like band gaps.
\end{abstract}

\maketitle

\paragraph{\textcolor{blue}{Introduction.}}
Periodic band structures and spin-orbit coupling play a key role in many modern condensed matter contexts. Implementing these aspects with ultracold quantum gases, using 
optical lattices~\cite{Denschlag2002, Morsch2006, Dan2012} and 
Raman dressing~\cite{Lin2011, Higbie2004} has opened up avenues for research providing deep insights into topics including 
synthetic gauge fields~\cite{Engels2015Feb, Gadway2018, Huang2016}, 
optical flux lattices~\cite{Cooper2011Apr}, 
topological state spaces~\cite{Konotop2018, Weifeng2019}, 
supersolids~\cite{Pan2012, Engels2019May, WKetterle2017, ShouGang2018}, 
and more~\cite{Wu2012, Kouwenhoven2010, Gong2019Nov, ChuanweiXu, Stringari2016}. 
The effective Hamiltonian associated with spin-orbit coupling (SOC) is distinct from that of an optical lattice in an important way: spin-orbit coupling produces a double-well structure in momentum space that is accompanied by a corresponding change of the spin composition of the states. Since the states coupled by the Raman beams are in different spin states, they do not interfere with each other, and no periodic lattice structure is induced. 

Supplementing the spin-orbit coupling with a suitably chosen radio frequency (RF) drive leads to the emergence of an effective lattice structure even though neither the spin-orbit coupling nor the radio frequency alone produce a periodic structure. In an intuitive picture, the RF drive mixes the spin states that are coupled by the Raman drive, while maintaining their momentum states. Then the two coupled states, separated in momentum space by the Raman momentum, can interfere, leading to a periodic structure. This synthesized spin-orbit-coupled lattice is called the Zeeman lattice. A static version has been introduced in early works by~\cite{Spielman2012} and \cite{Zwierlein2012Aug} for bosonic and fermionic systems, respectively, to generate artificial gauge fields in the presence of periodic potentials. 

Here, we investigate dynamics in the Zeeman lattice and demonstrate a variety of lattice manipulation techniques available to this system. Due to Galilean invariance, accelerating the atoms by an external force can be substituted by appropriate changes of the Raman drive frequency. This opens the door to robust manipulation schemes. We demonstrate the existence of the Zeeman lattice using Kapitza-Dirac scattering, study Bloch oscillations via an acceleration of the Zeeman lattice, and spectroscopically probe inter-band transitions by shaking the lattice resonantly. This work shows that Zeeman lattices provide a flexible and robust system for future studies of spin-selective lattices~\cite{Aidelsburger2013, WKetterle2013}, Floquet engineering~\cite{Jing2018}, and the study of materials properties such as flat band physics~\cite{Chuanwei2013} and semimetal-like band structures.

\begin{figure}
\centering
\includegraphics[scale=1.0]{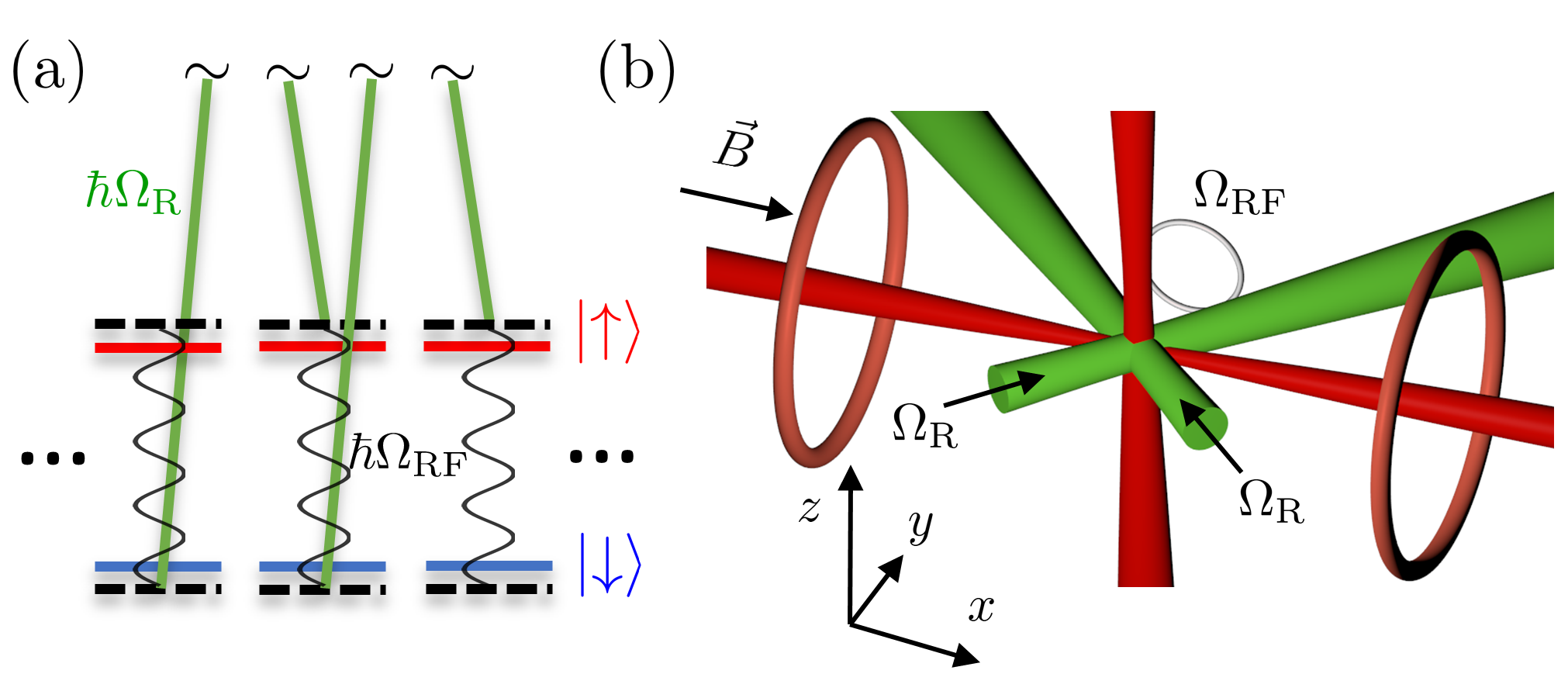}
\caption{(a) Coupling scheme employed to generate the Zeeman lattice. The horizontal direction indicates momentum, with a splitting of two Raman momenta between adjacent states. (b) Schematic of the experimental configuration. Two Raman beams are incident on the center of a crossed optical dipole trap at approximately 45$^\circ$ angles relative to the $x$-axis.}
\label{fig-exp setup}
\end{figure}

\paragraph{\textcolor{blue}{Theoretical Framework.}}\label{sec:theory}
The Zeeman lattice is realized via the simultaneous coupling of 
two hyperfine states using a pair of Raman lasers and an external RF field. The couplings produced by the Raman and RF drive are schematically demonstrated in the bare (uncoupled) basis in Fig.~\ref{fig-exp setup}(a). 
The two coupled hyperfine states are considered as the two spin orientations of a pseudo-spin 1/2 system. The Raman beams (marked $\Omega_\R$ in Fig.~\ref{fig-exp setup}(b)) are arranged such that they induce spin-orbit coupling by imparting momentum on the atoms while also flipping the spins in a two-photon transition. The RF coupling flips the spins without changing the momentum state. 

This coupling scheme is described by the Hamiltonian
\begin{equation}
 H_0= \frac{p_x^2}{2m} + \begin{bmatrix}-\Delta\epsilon/2
 &\mathcal{C} \\
\mathcal{C}^* & \Delta\epsilon/2
\end{bmatrix}, \label{eq:Hamiltonian}
\end{equation}
where
$\mathcal{C} = \frac{\Omega_\text{R}}{2}  
\exp \left[i \left(2k_{\text{R}} x + \varphi(t)\right) \right] +  \frac{\Omega_\text{RF}}{2} \exp\left(i\omega_{\text{RF}} t \right)$. $\Omega_{\text{R}}$ is the Rabi frequency due to the Raman coupling, $\Omega_{\text{RF}}$ is the RF Rabi frequency, and 
$\omega_{\text{RF}}$ is the angular frequency of the  RF field. 
The Raman coupling involves a momentum exchange of $2\hbar k_{\text{R}} $ between the atoms and the Raman beams, where $ k_{\text{R}} $ is the wave vector of Raman lasers. 
The energy associated with the two-photon recoil is the recoil energy given by $E_\mathrm{R}=\hbar^2k_\R^2/2m$.
The phase $\varphi(t)$ is proportional to the angular frequency difference $\Delta\omega_{\text{R}}$ between the Raman lasers, which can readily be tuned in the experiment.
In the Hamiltonian $H_0$, $m$ is the atomic mass and
$\Delta\epsilon$ is the energy difference between the two hyperfine states.

In order to show the Zeeman lattice explicitly, the Hamiltonian in a co-moving frame with the Raman field is obtained by the unitary transformation 
$U = \exp \left[ i \left( 2k_{\text{R}} x + \varphi(t) \right)  \sigma_z/2 \right] $ 
to the Hamiltonian in Eq.~\ref{eq:Hamiltonian}, leading to  $H=U^\dagger H_0 U-i\hbar U^\dagger \partial U /\partial t$ with
\begin{align}
H = \frac{p^2_x}{2m}  + \frac{1}{2} \left( \mathbf{B}_\text{latt}(x)  + \mathbf{B}_\text{soc} \right) \cdot \mathbf{S}
\label{eq:H}
\end{align}
Here, $\mathbf{S}=\left( \sigma_x,\sigma_y,\sigma_z\right)$ are the Pauli matrices. The Zeeman lattice is denoted by $\mathbf{B}_\text{latt}(x) =(b_x,b_y,b_z)$ with 
$b_x= \hbar\Omega_\text{R}+\hbar\Omega_\text{RF} 
\cos\left[2k_\text{R}\left( x + x_0(t) \right)\right]$, 
$b_y=\hbar\Omega_\text{RF} 
\sin\left[ 2k_\text{R} \left( x + x_0(t) \right) \right]$, 
the detuning $b_z= \hbar  d \varphi/d t -  \Delta\epsilon$,
and the time-dependent position offset $x_0(t)=(\varphi(t)-\omega_\text{RF}t)/2k_\text{R}$.
Therefore, this spin-orbit-coupled lattice is a Zeeman lattice with the velocity 
\begin{equation}
v_\text{lat} = \frac{ d \varphi/dt  - \omega_\text{RF}}{2k_\text{R}}. \label{eq:lat_vel}
\end{equation}
The spin-orbit coupling is represented by $ \mathbf{B}_\text{soc} = \left( 0, 0, 2\hbar k_\text{R} p_x/m\right) $.  The Zeeman lattice becomes stationary when $\varphi(t)=\omega_\text{RF}t$ which implies $b_z= \hbar  \omega_\text{RF}-  \Delta\epsilon$. The stationary Zeeman lattice can be implemented by tuning the angular frequency difference  $\Delta\omega_{\text{R}}$ between the Raman lasers to exactly equal $ \omega_\text{RF}$. 

\begin{figure}
	\centering
	\includegraphics[scale=.9]{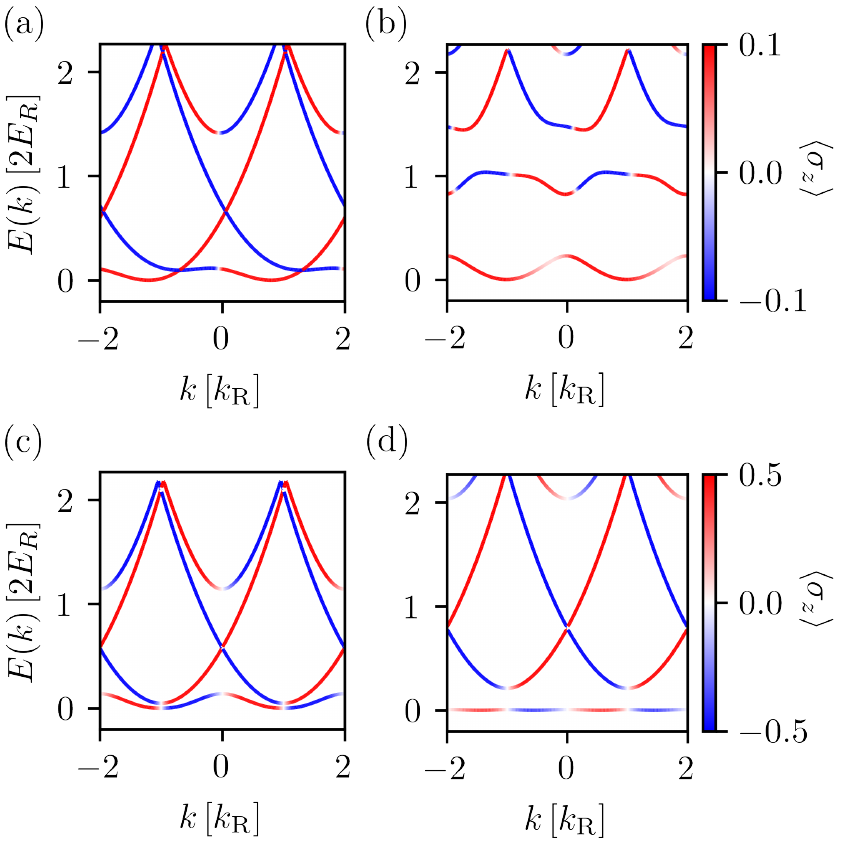}
	\caption{Band structure over two Brillouin zones where the plot colors indicate the spin polarization $\langle\sigma_z\rangle$.
		(a) Band structure with spin-orbit coupling in the absence of RF coupling. The parameters are $\hbar\Omega_\R = 2.6\,E_\R$, $\hbar\Omega_\mathrm{RF} = 0$  and the detuning $b_z = 2\pi \times 500$\, Hz.  
		(b) Band structure of the stationary Zeeman lattice as described in Eq.~\ref{eq:H}. The values used in this figure are $\hbar\Omega_\R = 2.6\,E_\R$, $\hbar\Omega_\mathrm{RF}=2.3\,E_\R$ and the detuning 
		$b_z = 2\pi \times 500$\,Hz. 
		(c) Semimetal-like band structure in Zeeman lattice. 
		The values used in this figure are $\hbar\Omega_\R = 2.0\,E_\R$, $\hbar\Omega_\mathrm{RF} = 0.10\,E_\R$ and the detuning $b_z = 0$\,Hz.
		(d) Flat band in Zeeman lattice. The values used in this figure are $\hbar\Omega_\R = 4.0\,E_\R$, $\hbar\Omega_\mathrm{RF} = 0.48\,E_\R$ and the detuning $b_z = 0$\,Hz.}
	\label{fig-band structure}
\end{figure}

The lattice structure of the above spin-orbit-RF coupled Hamiltonian (Eq.~\ref{eq:H}) can be discussed from several different perspectives. Fig.~\ref{fig-band structure} shows the band structure of the stationary Zeeman lattice in various parameter regimes where the spin polarization
$\langle\sigma_z\rangle$ is indicated by the color of the curve and given by $\langle\sigma_z\rangle = (N_\uparrow-N_\downarrow)/N_\mathrm{tot}$. Here, $N_\uparrow$ and $N_\downarrow$ are the occupation of 
the spin up $\lvert \uparrow \rangle$ and spin down $\lvert \downarrow \rangle$ state, respectively, and $N_\mathrm{tot} = N_\uparrow + N_\downarrow$.
If one applies Bloch's theorem to diagonalize the spin-orbit coupling Hamiltonian (without the RF coupling) assuming an infinite set of plane wave solutions, one finds that bands in different Brillouin zones are not coupled and therefore are trivially inaccessible, shown in Fig.~\ref{fig-band structure}(a). The inclusion of the RF coupling then opens up gaps and produces a periodic band structure (cf. Fig.~\ref{fig-band structure}(b)). 

The Hamiltonian given above can be tuned in highly flexible ways.
For example, it can result in a semimetal-like band structure (Fig.~\ref{fig-band structure}(c)) or in a flat band as the lowest Bloch band (Fig.~\ref{fig-band structure}(d)) in different parameter regimes. 

A theoretical analysis~\cite{[{See Supplemental Material}][{ for the details on the Galilean transform of the Zeeman lattice.}]supp} reveals that applying an external force to the atoms in a stationary Zeeman lattice is equivalent to accelerating the Zeeman lattice without an external force when viewed from the non-inertial reference frame of the lattice. Furthermore, an arbitrary acceleration of the lattice potential leaves the Zeeman lattice Hamiltonian invariant up to an additional term which represents the apparent force due to the motion of the noninertial reference frame given by 
$\mathcal{F}= - (m/2k_\text{R}) \, d^2\varphi/dt^2$. This force term appears as a detuning in the lattice Hamiltonian, further showing the equivalence between lattice acceleration and frequency detuning between the two Raman beams. 

The analysis in~\cite{supp} also reveals that the Zeeman lattice velocity $v_\text{lat}$ (Eq.~\ref{eq:lat_vel}) depends on both the angular frequency $\omega_{\text{RF}}$ of the RF field and the Raman detuning $\Delta\omega_{\text{R}}$, while the effective detuning $b_z$ and the lattice amplitude are independent of $\Delta\omega_{\text{R}}$. Therefore, in our experiments described below we only change $\Delta\omega_{\text{R}}$ and keep $\omega_{\text{RF}}$ constant. 

\paragraph{\textcolor{blue}{Experimental Setup.}}\label{sec:setup}
In our experiment, we prepare a BEC of approximately $2.5\times 10^5$ $^{87}$Rb atoms in the $\lvert F, m_F\rangle = \lvert 1, -1\rangle$ hyperfine state. The BEC is confined in a crossed optical dipole trap characterized by the harmonic trap frequencies $\bm\omega = (\omega_x, \omega_y , \omega_z) = 2\pi(20, 160, 190)$ Hz. A 10 G bias field applied along the $x$-axis lifts the degeneracy among the three states in the $F=1$ hyperfine manifold.
We generate  spin-orbit coupling in the $x$-direction using two Raman beams with a wavelength of  $789$~nm intersecting at approximately $45^o$ angles with the $x$-axis, as shown in Fig.~\ref{fig-exp setup}(b).
This produces an effective recoil energy of $E_\text{R} = h \times 1.9$ kHz. The frequencies of the Raman beams are tuned so that the $| 1, -1\rangle$ and $| 1, 0\rangle$ hyperfine states are nearly resonantly coupled via the two-photon Raman transition. These two states are additionally coupled by an RF drive (Fig.~\ref{fig-exp setup}(a)). 
The quadratic Zeeman shift places the $\lvert 1, +1\rangle$ state far out of resonance so that effectively a spin-$1/2$ system composed of 
$\lvert\uparrow \rangle \equiv | 1,-1\rangle$ and $\lvert\downarrow \rangle \equiv | 1, 0 \rangle$ as the two pseudo-spin states is realized.  
After performing the experiments described in the following sections, imaging is performed by suddenly turning off the trap and all driving fields, and allowing time of flight in the presence of a Stern-Gerlach field. An absorption image is then taken which resolves the BEC into the bare-state spin and momentum components. 

\paragraph{\textcolor{blue}{Kapitza-Dirac Scattering.}}\label{sec:KDscattering}
To experimentally demonstrate the presence of a lattice structure, Kapitza-Dirac (KD) scattering can be employed. KD scattering is generated by abruptly applying a lattice potential for a brief amount of time. In order to demonstrate KD scattering, we begin by adiabatically dressing the BEC with the RF field:
we prepare a balanced mixture of the $|1,-1\rangle$ and $|1,0\rangle$ states by ramping up the intensity of a far detuned ($100$ kHz) RF field to a Rabi coupling strength of $8.6 \, E_\mathrm{R}$ and then adiabatically reducing the detuning down to zero. 

\begin{figure}
\centering
\includegraphics[scale=1]{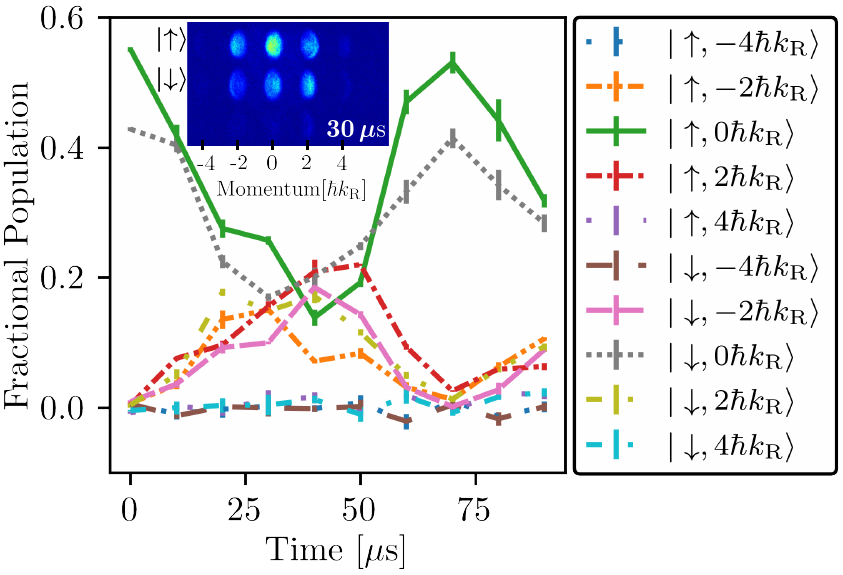}
\caption{Kapitza-Dirac scattering due to sudden application of the Zeeman lattice. Fractional state populations, normalized to the total atom number, are plotted as a function of evolution time. 
The absorption image shows the populations of various spin-momentum states at the evolution time of $30\,\mu$s.} 
\label{fig:KDresults}
\end{figure}

We then suddenly turn on the spin-orbit coupling where the frequency difference between the two Raman beams is set equal to the RF frequency and the Raman coupling strength is equal to the RF coupling strength. This suddenly establishes a stationary effective lattice potential and leads to momentum space oscillations associated with KD scattering. 

We vary the length of this pulse from $0 \, \mu \text{s}$ to $90 \, \mu \text{s}$ to observe the dynamical evolution of the system in both the momentum and spin basis. Fig.~\ref{fig:KDresults} shows the time evolution of the momentum states after the emergent Zeeman lattice has been introduced. 
The occupation of the $\lvert\uparrow,-2\hbar k_\R \rangle$ and $\lvert\downarrow,2\hbar k_\R \rangle$ states clearly distinguishes the lattice-like coupling from spin-orbit coupling without RF coupling: these two states would not be populated in a purely spin-orbit coupled system. They require the presence of both the Raman drive and the RF drive.
We observe approximately symmetric dynamics in the momentum populations consistent with the abrupt application of a stationary lattice potential, as expected from our choice of $\delta\omega_\mathrm{R} = \omega_\mathrm{RF}$ in this experiment.  
Furthermore, the congruent evolution of both spin populations indicates that the RF coupling is strong enough to maintain the lattice structure.

\paragraph{\textcolor{blue}{Bloch oscillation in the Zeeman lattice.}}\label{sec:BlochOsc}
A hallmark feature of periodic potentials is the possibility to observe Bloch oscillations, i.e. the oscillatory motion of particles under the influence of a constant force. Here we explore the band structure shown in Fig.~\ref{fig-band structure}(b) where the lowest band is well isolated from the first excited band, suppressing Landau-Zener tunneling \cite{Peik1996}.

\begin{figure}
\centering
\includegraphics{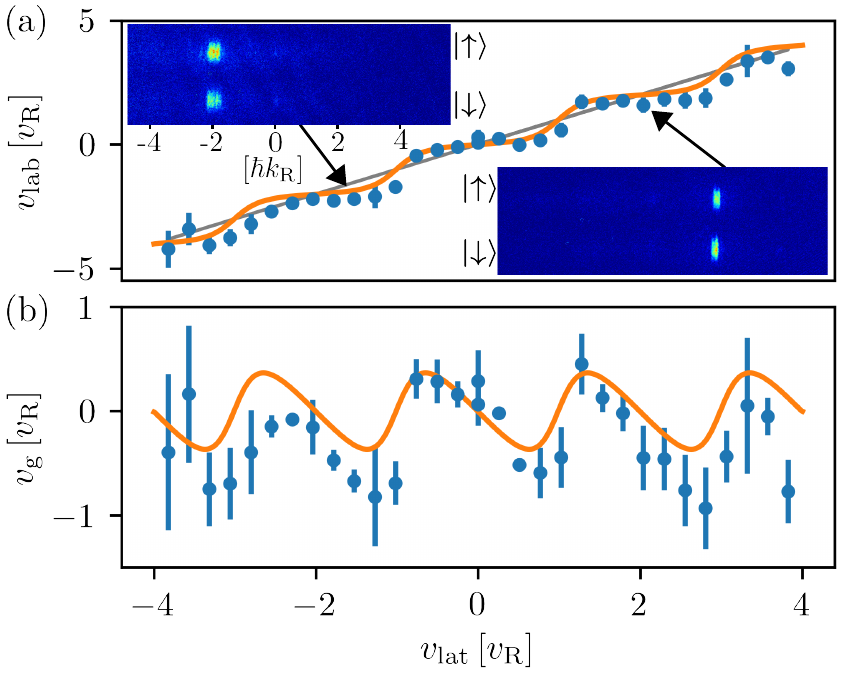}
\caption{
(a) Velocity in the lab frame versus lattice velocity. The absorption images are given at two different instances where atoms occupy different spin-momentum states due to Bloch oscillation. The underlying straight line corresponds to 
$v_\text{lab} = v_\text{lat}$. (b) Group velocity versus lattice velocity. The blue dots represent the average of three experimental shots and the orange line represents the theoretical results based on the curvature of the ground state band. The velocities are reported in units of recoil velocity, $v_\R =\hbar k_\R/m$, but can equivalently be read as momentum in units of $k_\R$. The system was prepared with $\hbar\Omega_\R = 2.6\,E_\text{R}$, $\hbar\Omega_\text{RF} = 2.3\,E_\text{R}$, $b_z = 2\pi \times 500$ Hz.
}
\label{fig:BO_exp}
\end{figure}

We begin by preparing the BEC in a minimum of the lowest band of the Zeeman lattice: we adiabatically dress the condensate with a Raman coupling field of strength $\hbar\Omega_\text{R} = 2.6\, E_\text{R}$ and a detuning $b_z = 2\pi \times 500$ Hz, and then adiabatically ramp on the strength of the radio frequency drive  to a coupling strength of 
$\hbar\Omega_\text{RF} = 2.3\,E_\text{R}$ over $100$~ms while satisfying the stationary lattice condition $\varphi-\omega_\text{RF}t =0$. 
To experimentally demonstrate Bloch oscillations in the system, we then ramp the frequency difference between the Raman beams in such a way that the relative phase between the two Raman beams evolves as $ \varphi(t)= \omega_\text{RF}t + \alpha t^2$ with 
$\alpha = +15$ MHz/s, or, in different experimental runs, $\alpha = - 15$ MHz/s.
Therefore the lattice is accelerated as $x_0(t)=(\varphi(t)-\omega_\text{RF}t)/2k_\text{R}=\alpha t^2/2k_\text{R}$ with the constant 
acceleration $\alpha/k_\text{R} = + 2.6\ \text{m}/\text{s}^2$, or 
$\alpha/k_\text{R} = - 2.6\ \text{m}/\text{s}^2$, respectively. 
The velocity of the lattice evolves as $v_\text{lat}=\alpha t/k_\text{R} $. 
Equivalently, in the frame of the moving lattice atoms feel a constant force $\mathcal{F}=-m\alpha/k_\text{R}$~\cite{supp}.
It linearly changes  the quasimomentum of the atoms, $k(t)=-k_\text{R}+\mathcal{F} t/\hbar= -k_\text{R}-mv_\text{lat}/ \hbar $. The acceleration is applied for up to 2 ms, which corresponds to moving through more than two Brillouin zones.  The ramp can be performed in either direction (i.e., $\pm \alpha$), yielding four Brillouin zones worth of data.  At various stages along the ramp, and thus for various lattice velocities, we perform expansion imaging of the 
BEC and from the observed velocity components determine the average lab-frame velocity $v_\text{lab}$ of the cloud. The obtained $v_\text{lab}$  as a function of $v_\text{lat}$  are presented in Fig.~\ref{fig:BO_exp}(a). The staircase structure is the characteristics of Bloch oscillations. In the co-moving frame with the lattice, the lattice is at rest and the BEC has the group velocity, $v_g = v_\text{lab} - v_\text{lat}$, which forms the sinusoidal pattern as shown in 
Fig.~\ref{fig:BO_exp}(b).  In theory, the group velocity can be predicted by calculating the gradient of the lowest band $\epsilon_0 (k)$ shown in Fig.~\ref{fig-band structure}(b), $v_g =  \partial \epsilon_0 (k) / \hbar \partial k $. The theoretically calculated group velocity is shown by the orange curves in Fig.~\ref{fig:BO_exp}. The good agreement between the observations and theoretical predictions confirms that we can effectively accelerate atoms in the spin-orbit-coupled lattice by accelerating the lattice instead. 

\paragraph{\textcolor{blue}{Band Spectroscopy of the Zeeman lattice.}}\label{sec:Shaking}

For our demonstration of Bloch oscillations, a constant force on the atoms has effectively been implemented by ramping the frequency difference between the Raman beams. In the following, we show that a periodic modulation of this frequency difference leads to a shaken Zeeman lattice which can induce inter-band transitions. This provides a practical means for performing band spectroscopy in the Zeeman lattice.

\begin{figure}
\centering
\includegraphics{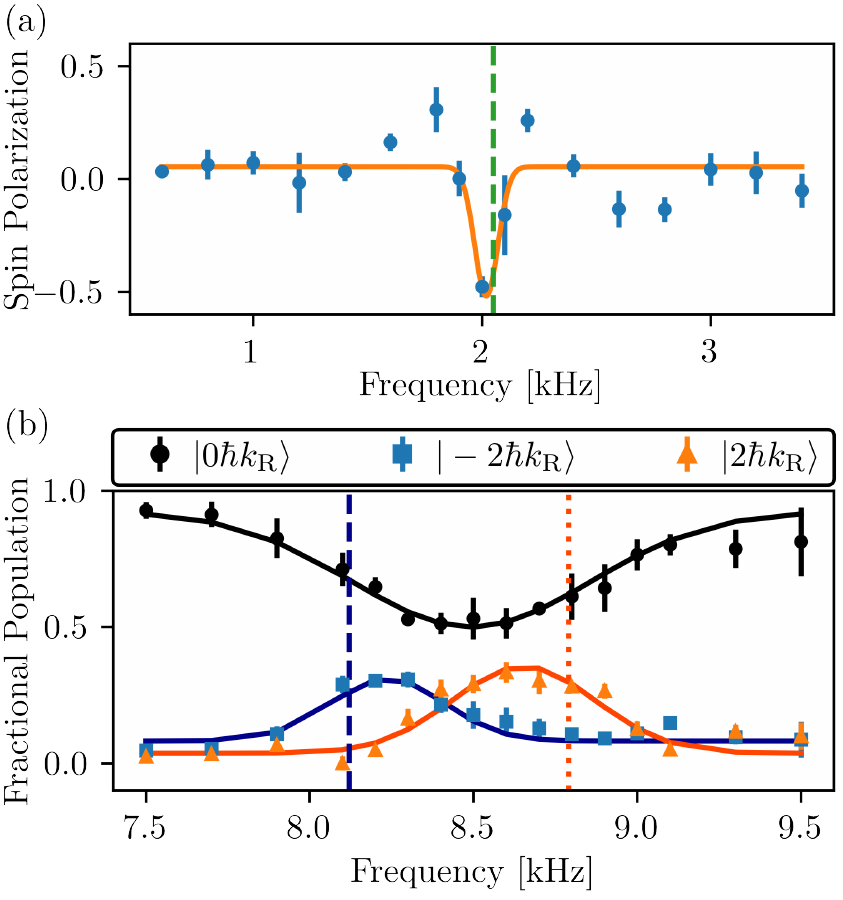}
\caption{(a) Lattice band spectroscopy measuring the spin-polarization as a function of lattice shaking frequency near the $n=0\rightarrow 1$ resonance. The resonance is indicated by a strong spin polarization response. The vertical dashed line corresponds to the predicted resonance frequency calculated for the $n=0\rightarrow 1$ transition using an infinite, noninteracting, two-state model.
(b) Lattice band spectroscopy measuring the momentum-space response as a function of lattice shaking frequency near the $n=0\rightarrow 2, 3$ resonances. The vertical dashed and dotted lines correspond to the predicted resonance frequency for the $n = 0 \rightarrow 2$ transitions and $n = 0 \rightarrow 3$ transition, respectively.}
\label{fig-band exc}
\end{figure}

Here, we begin with a BEC prepared in the ground state of the stationary Zeeman lattice with
$\hbar\Omega_R = 2.6\,E_\text{R}$, $\hbar\Omega_\text{RF} = 1.15\,E_\text{R}$, and $b_z = 2\pi \times 500$ Hz.
To modulate the phase of the lattice~\cite{Denschlag2002} we apply a sinusoidal modulation to the Raman frequency difference such that  $x_0(t) = (\varphi(t)-\omega_\text{RF}t)/2k_\text{R}=\phi_0\sin( 2\pi f t) /2k_\text{R}$, where $\phi_0/2k_\text{R}$ and $f$ are the shaking amplitude and frequency respectively. In the co-moving frame with the shaken lattice, atoms experience the oscillating force as 
$\mathcal{F}= 2 \pi^2 m \phi_0 f^2 \sin(2 \pi ft)/ k_\text{R}$. To induce inter-band transition,  the amplitude of  $\mathcal{F}$ should be a perturbation to the system. We shake the lattice with an amplitude of $\phi_0/2k_\text{R} = \pi/20k_\text{R}$ for 6 ms using various frequencies $f$. The measured spin polarization as a function of $f$ is shown in Fig.~\ref{fig-band exc} (a). The observed response of the spin polarization centered at $f = (2.02 \pm 0.05)$ kHz is the signature of an inter-band transition. 

Further transitions can be observed by driving the system at higher frequencies. Here, we modulate the phase by $\phi_0/2k_\text{R} = \pi/40k_\text{R}$ for 2 ms at each frequency. 
In this case, the clearest signature of the transition is the fractional population in the bare momentum states, which can be understood by inspecting the spin and bare-momentum composition of the underlying bands. The availability of different but correlated observables for the spectroscopy is a particular strength of our Zeeman lattice.
Fig.~\ref{fig-band exc}(b) shows that the fractional population of the zero-momentum state exhibits a single broad feature, indicating transitions out of the ground band. Plotting the fractional populations in the $\pm2k_\text{R}$ states resolves this broad peak into two peaks centered at $(8.24 \pm 0.04)$ kHz and $(8.65 \pm 0.02)$ kHz, respectively, where the uncertainty in the line center is given by the standard error of the fit to the experimental data. 

These observed features are in reasonable agreement with the theory calculations for the excitations to the $n = 1, 2, 3$ states. For a BEC placed at a minimum of the lowest band, the diagonalization of Eq.~\ref{eq:H} for the stationary lattice predicts the $n = 0 \rightarrow 1,2,3$ transitions to occur at frequencies $2.05$ kHz, $8.12$ kHz, and $8.79$ kHz for our experimental parameters, as marked with dashed lines in Fig.~\ref{fig-band exc}. These values are calculated based on a two-state model for an infinite, noninteracting, two-state system. We note that the resonance frequencies are quite sensitive to the atomic momentum -- displacing the momentum along the noninteracting dispersion by just $0.02 \, k_\mathrm{R}$ produces a near perfect agreement between experiment and theory for all three resonances observed here.

\paragraph{\textcolor{blue}{Conclusion.}}

In conclusion, we have studied the rich dynamics of atoms in a Zeeman lattice that emerges from the confluence of spin-orbit coupling and an external radio frequency field. 
We have demonstrated an extensive set of experimental techniques to characterize the Zeeman lattice including KD scattering, Bloch oscillation, and resonant band spectroscopy.
In our band spectroscopy, we have shown that that both spin and momentum composition are useful observables, demonstrating the multifaceted nature of the Zeeman lattice. 
The availability of a variety of correlated observables, such as spin and momentum populations, affords powerful experimental tools for performing detailed studies as we have shown with the band spectroscopy. 
This work opens the door to a range of further investigations including spin-dependent lattices, the characterization of a semimetal-like band structure 
(Fig.~\ref{fig-band structure}(c)), dynamics of a BEC in a tunable flat ground band (Fig.~\ref{fig-band structure}(d)), the realization of Wannier-Stark ladders~\cite{Raizen1996}, and exploration of the spin Hall effect~\cite{Sayan2018}, to name a few.

\paragraph{\textcolor{blue}{Acknowledgement.}}
M.K.H.O., A.M., E.C., S.M., T.B., and P.E. acknowledge funding from NSF through Grant No. PHY-1912540. H.H. and Y.Z. are supported by National Natural Science Foundation of China with Grants Nos.~11974235 and 11774219.



%


\begin{widetext}
	
\section{Supplemental Material}
For the study of lattice dynamics, it is instructive to discuss the role of Galilean invariance in the system. For this, we first consider a system in which an optical lattice and spin-orbit coupling are realized independently by two different sets of lasers. This is described by the Hamiltonian~\cite{Engels2015Feb}
$$H_0'= \frac{p_x^2}{2m} + V\sin\left[2k_\text{latt}(x+x_0(t) )\right] + \frac{\hbar k_\text{R} p_x}{m}\sigma_z.$$ 
The optical lattice is given by  $V\sin\left[2k_\text{latt}(x+x_0(t) )\right]$ with lattice depth $V$, lattice laser wave vector $k_\text{latt}$ and time-dependent position offset $x_0(t)$, all of which can be precisely manipulated in experiment.   This  spin-orbit-coupled optical lattice breaks Galilean invariance in the following sense: The Galilean transformation is 
\begin{equation*}
		G = \exp\left[\frac{im}{2\hbar}\int_0^t \left( \frac{dx_0(\tau)}{d\tau} \right)^2 d\tau \right]
		\exp\left[\frac{-im \dot{x_0}(t) x}{\hbar} \right] 
		 \exp\left[ \frac{i x_0(t)p_x}{\hbar} \right]
\end{equation*}
with $G^\dagger x G=x-x_0(t)$ and $G^\dagger p_x G=p_x-mdx_0/dt$~\cite{Messiah}. It transforms the system into the co-moving frame with the optical lattice, 
\begin{equation*}
		G^\dagger H_0' G - i\hbar G^\dagger \frac{\partial G}{\partial t} = \frac{p_x^2}{2m} + V\sin\left(2k_\text{latt}x \right) 
		 + \frac{\hbar k_\text{R}}{m} \left(p_x - m \frac{dx_0}{dt}\right) \, \sigma_z - mx \frac{d^2x_0}{dt^2}. 
\end{equation*}
The breakdown of Galilean invariance is indicated by the appearance of the term $\hbar k_\text{R}\sigma_zdx_0/dt$. Physically, the optical lattice lasers and spin-orbit coupling Raman lasers build up two independent time-dependent frames. The changing of the  frame of optical lattice lasers inevitably affects the frame of Raman lasers.

For the Zeeman lattice in 
Eq.~\ref{eq:H}, 
the time-dependence relates to both Raman lasers and the RF field, i.e., $x_0(t)=(\varphi(t)-\omega_\text{RF}t)/2k_\text{R}$. Transforming into this joint co-moving frame by applying the Galilean transformation generates 
$H_\text{com}=G^\dagger HG-i\hbar G^\dagger \partial G/\partial t$,
\begin{equation}
	H_\text{com}=\frac{p^2_x}{2m}  +\frac{1}{2} \left( \mathbf{B}'_\text{latt}(x)  + \mathbf{B}_\text{soc} \right) \cdot \mathbf{S} +\mathcal{F}x.
	\label{comoving}
\end{equation}
Here $\mathbf{B}'_\text{latt}(x) =(b'_x,b'_y,b'_z)$ with 
$b'_x= \hbar\Omega_\text{R} + \hbar\Omega_\text{RF} \cos\left(2k_\text{R}x\right)$, $b'_y=\hbar\Omega_\text{RF} \sin\left(2k_\text{R}x\right)$,  
$b'_z= \hbar\omega_\text{RF}  -  \Delta\epsilon$, and 
$\mathcal{F}=-m d^2x_0 /dt^2$. 
Note that $\mathbf{B}'_\text{latt}(x)$ is exactly the same as for the stationary Zeeman lattice ($\varphi(t)=\omega_\text{RF}t$) in 
Eq.~\ref{eq:H},
which indicates that the Zeeman lattice restores Galilean invariance. By comparing Eq.~\ref{comoving} with the Hamiltonian 
$H$ in 
Eq.~\ref{eq:H}, 
we see that Eq.~\ref{comoving} is precisely the Hamiltonian for an atomic cloud moving with velocity $v=(\Delta \omega_\text{R}-\omega_\text{RF})/2k_\text{R}$ in the stationary Zeeman lattice. This equivalence between stationary atoms in a moving lattice and moving atoms in a stationary lattice is taken as a signature of the restored Galilean invariance in the system. 

The Raman beams and RF field together form a joint time-dependent frame. Co-moving with this frame, atoms equivalently feel the force $\mathcal{F}=-m d^2x_0 /dt^2 = - (m/2k_\text{R}) \, d^2\varphi/dt^2$ so that their external degree of freedom can be manipulated without changing the internal degree. From an experimental point of view, this is an important aspect: moving, accelerating or shaking a lattice by changing the laser frequency in one of the beams can typically be performed with much higher precision than directly moving the atoms. We note that also the independent implementation of spin-orbit coupling and an optical lattice with two separate sets of lasers affords this experimental possibility, but then both sets of lasers need to be frequency tuned in the same way. In case of the Zeeman lattice, the RF coupling does not produce an additional reference frame for the atoms due to the negligible Doppler shift associated with RF frequencies. The Zeeman lattice presented here produces a spin-dependent lattice with a reference frame which only depends on one parameter -- the Raman detuning. 

\end{widetext}

\end{document}